\documentclass[aps,prl,reprint,superscriptaddress,nofootinbib,floatfix]{revtex4-2}
\usepackage{amsmath,amssymb,bm,graphicx,xcolor}
\graphicspath{{figures_revised/}{figures/}}
\begin{document}

\title{Charge-Selective Displacement of Spatiotemporal Optical Vortices at a Spatially Uniform Time Boundary}
\author{Chao Mei}
\email{meichao@nbu.edu.cn}
\affiliation{Department of Physics, School of Physical Science and Technology, Ningbo University, Ningbo 315211, China}

\author{Peilong Yang}
\email{yangpeilong@nbu.edu.cn}
\affiliation{Laboratory of Infrared Materials and Devices,
    The Research Institute of Advanced Technologies, Ningbo University, Ningbo 315211, China}

\begin{abstract}
At a spatially uniform, in-plane-isotropic time boundary, the transverse displacement of a spatiotemporal optical vortex changes sign with its topological charge, despite a charge-independent conversion weight. Radial spectral filtering couples to the vortex phase while preserving identical converted spectral intensities for opposite charges. For a narrow scalar Gaussian spectrum, the displacement follows the logarithmic slope of the spectral weight and occurs with zero mean transverse momentum. In a finite pumped GaAs waveguide, unequal attenuation during extraction adds angular deflection. Independent Maxwell--Drude simulations test this extraction-induced response and give $\pm42$~mrad before detector filtering. Calculated centroid traces at multiple planes separate propagation-induced drift from position offsets.
\end{abstract}
\maketitle

\textit{Introduction.—}Opposite vortex charges can be indistinguishable in spectral intensity while carrying opposite phase circulation. In spatiotemporal optical vortices (STOVs), the phase winds in a mixed space--time plane and the orbital angular momentum (OAM) is transverse~\cite{Jhajj2016,Chong2020,Hancock2021}. A transverse displacement would expose the hidden handedness as a difference in position between separately prepared $\ell=+1$ and $\ell=-1$ packets. The central challenge is to obtain the displacement from a boundary with no transverse directional preference. Transverse STOV shifts at spatial interfaces depend on Fresnel geometry~\cite{Mazanov2022,Tang2024}. Reflection from a relativistic flying mirror introduces boundary motion~\cite{Wu2026}. A spatially uniform time boundary changes frequency while conserving wave vector~\cite{Mendonca2005,Moussa2023}. In-plane isotropy further gives identical conversion coefficients for mirror-related transverse wave vectors. Opposite vortex charges consequently retain identical converted spectral intensities. The temporal Goos--H\"anchen effect describes a longitudinal delay~\cite{Ponomarenko2022}. The question is whether the vortex phase and a common radial conversion weight can produce opposite spatial centroids while preserving identical output spectral intensities.

Radial spectral filtering weights the opposite phase gradients of the two vortices and produces opposite centroid shifts. For the scalar Gaussian STOV spectrum below, the exact centroid relation reduces in the narrowband limit to a logarithmic-slope law. This boundary-induced shift leaves the mean transverse wave number zero. In a finite pumped GaAs waveguide, unequal attenuation during extraction supplies transverse momentum and adds deflection. Independent Maxwell--Drude simulations test this extraction-induced angular response. Measurements at multiple propagation distances separate the drift from the position offset of the extracted record. Recovering the displacement at the time boundary additionally requires accounting for extraction and the detector response.

\textit{General displacement law.—}STOV packets with opposite topological charges and identical spectral intensities are prepared and launched separately. Each packet encounters the same spatially uniform, in-plane-isotropic time boundary. The carrier propagates along $+z$, and $x$ is the transverse coordinate in the vortex plane. The scalar incident spectra are~\cite{Chong2020,Hancock2021}
\begin{equation}
F_\ell^{\rm in}(k_x,k_z)=\frac{\ell w_z(k_z-k_{z0})-i w_xk_x}{2}\,G(k_x,k_z),
\label{eq:stov}
\end{equation}
where $\ell=\pm1$, $w_x$ and $w_z$ are the envelope widths conjugate to $k_x$ and $k_z$, and $G\ge0$ is an elliptical Gaussian centered at $(0,k_{z0})$ with $k_{z0}>0$.

\begin{figure}[!tbp]
\includegraphics[width=\columnwidth]{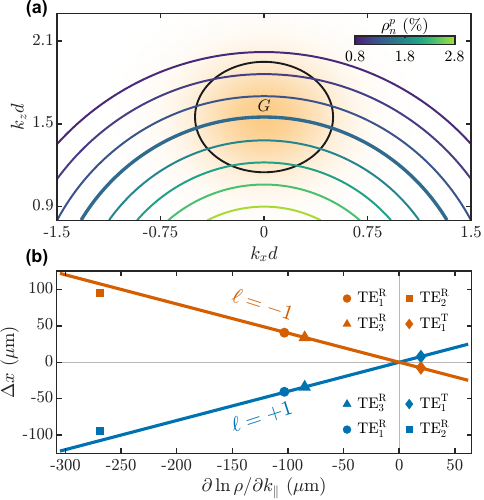}
\caption{Radial spectral filtering produces a transverse displacement. (a)~Contours of the time-reflected TE$_1$ spectral weight $\rho_1^{\rm R}$, colored by weight in percent. Orange shading shows the Gaussian spectral amplitude $G$. The black ellipse marks the constant-amplitude section $G=G_{\max}/e$, with semiaxes $2d/w_x$ and $2d/w_z$. (b)~Solid lines show the narrowband prediction of Eq.~(\ref{eq:slope}) for separately launched $\ell=\pm1$ packets. Symbols give Eq.~(\ref{eq:exact}) for the Maxwell--Drude time-reflected ($\mathrm{R}$) channels and the abrupt-boundary time-transmitted ($\mathrm{T}$) TE$_1$ channel. The GaAs example uses $d=100~\mu$m, $w_x=4d$, and $w_z=5d$. The derivative is evaluated at $k_{\parallel}=k_{z0}$. Reflected indices $n=2,3$ denote the TE$_2$- and TE$_3$-like radiative bands in Fig.~\ref{fig:theorem}.}
\label{fig:mechanism}
\end{figure}

For each output mode $n$ and temporal branch $p\in\{{\rm T},{\rm R}\}$, the boundary acts on the spectrum of each separately launched packet through a charge-independent complex coefficient $c_n^p(k_{\parallel})$. The symbol $k_{\parallel}=\sqrt{k_x^2+k_z^2}$ denotes the in-plane wave-number magnitude, and $\rho_n^p=|c_n^p|^2$ is the spectral intensity weight applied by the boundary. The branches ${\rm T}$ and ${\rm R}$ denote time transmission and reflection. Radial spectral filtering denotes weighting by $\rho_n^p(k_{\parallel})$, with no dependence on the in-plane wave-vector direction. For the centered spectrum in Eq.~(\ref{eq:stov}), the phase of $c_n^p(k_{\parallel})$ contributes equal and opposite terms at $k_x$ and $-k_x$, giving no net centroid shift. The phase gradient of the incident STOV still contributes to the spatial first moment. Its contribution gives the centroid
\begin{equation}
x_{n,\ell}^p=
\frac{\ell w_xw_z \int\rho_n^{p}(k_{\parallel})(k_z-k_{z0})G^2\,\mathrm{d}^2k}
{\int\rho_n^{p}(k_{\parallel})[w_x^2k_x^2+w_z^2(k_z-k_{z0})^2]G^2\,\mathrm{d}^2k}.
\label{eq:exact}
\end{equation}
Comparing the centroids from the two separate packets defines the charge-resolved half-difference $\Delta x_{n,\ell}^p=\ell[x_{n,+1}^p-x_{n,-1}^p]/2$. A common translation adds the same offset to both centroids and therefore cancels in the half-difference. For the centered incident spectrum, the half-difference equals the displacement induced by temporal conversion. For a sufficiently narrow spectrum, the radial geometry and spectral weight can be linearized about $(0,k_{z0})$. Provided $\rho_n^p(k_{z0})>0$, Eq.~(\ref{eq:exact}) then gives
\begin{equation}
\Delta x_{n,\ell}^p\simeq
\ell\frac{w_x}{2w_z}
\left.\frac{\partial\ln\rho_n^p}{\partial k_{\parallel}}\right|_{k_{z0}}.
\label{eq:slope}
\end{equation}
Near the spectral center $(0,k_{z0})$, the radial weight $\rho_n^p(k_{\parallel})$ varies predominantly along $k_z$ [Fig.~\ref{fig:mechanism}(a)]. The vortex phase converts the longitudinal spectral weighting into a transverse displacement. For $\ell=+1$, a negative local slope produces a shift toward $-x$, whereas Eq.~(\ref{eq:slope}) predicts zero displacement for a vanishing local slope [Fig.~\ref{fig:mechanism}(b)]. For the scalar input in Eq.~(\ref{eq:stov}), the slope law applies across material platforms whose conversion coefficient is radial and whose nonzero spectral weight can be linearized over the packet bandwidth. Equation~(\ref{eq:exact}) retains the full bandwidth dependence when the slope approximation fails. The output spectral intensity $\rho_n^p|F_\ell^{\rm in}|^2$ is even in $k_x$, so $\langle k_x\rangle=0$ and there is no angular deflection.

\begin{figure*}[!tp]
\includegraphics[width=\textwidth]{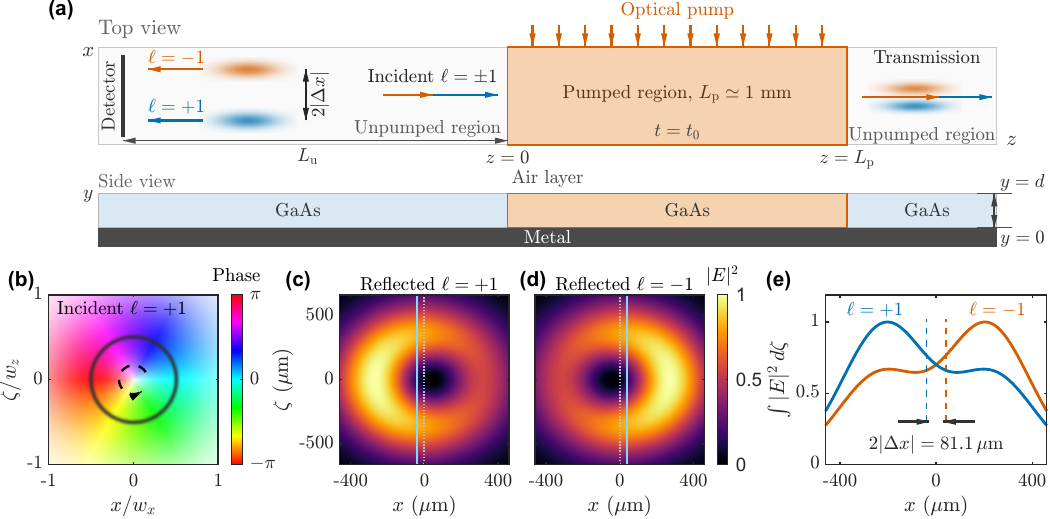}
\caption{GaAs realization of displacement and extraction. (a)~GaAs waveguide with a pumped region of length $L_{\rm p}$. Blue and orange denote $\ell=+1$ and $-1$, launched separately from the left. Time-reflected TE$_1$ packets exit at $z=0$ and are measured at distance $L_{\rm u}$ in the unpumped region. Transmitted packets exit at $z=L_{\rm p}$. (b)~Incident $\ell=+1$ packet in $(x,\zeta)$, where $\zeta=z-z_c$ measures position from the packet center $z_c=L_{\rm p}/2$ at the switch. Hue denotes phase, fading at low intensity. The gray ring marks the intensity maximum at $(x/w_x)^2+(\zeta/w_z)^2=1/2$, and the arrow indicates the phase circulation. The $\ell=-1$ packet follows from the $\ell=+1$ packet under $x\to-x$, up to a global phase. (c),(d)~Calculated reflected intensities at the reported experimental pump parameters. Each intensity map is normalized to its maximum. Dotted lines mark $x=0$ and cyan solid lines the centroids $\Delta x=\mp40.6~\mu$m. (e)~Intensities integrated over $\zeta$ and normalized to unit peak. Dashed lines mark the centroids from separate launches, separated by $2|\Delta x|=81.1~\mu$m.}
\label{fig:concept}
\end{figure*}

\textit{Time-boundary realization.—}Photoexcited GaAs waveguides provide an experimentally demonstrated platform for temporal conversion, with $8$--$23\%$ forward conversion at $0.35$--$0.48$~THz~\cite{Miyamaru2021,Takano2024}. The GaAs waveguide uses a slab of thickness $d=100~\mu$m, bounded by metal at $y=0$ and air above $y=d$ [Fig.~\ref{fig:concept}(a)]. The terahertz field is confined along $y$, and $n$ now labels transverse-electric (TE) modes. An optical pump at $y=d$ switches a region of length $L_{\rm p}$, with the incident packet centered at $z_c=L_{\rm p}/2$ [Fig.~\ref{fig:concept}(b)]. The conversion coefficient is calculated for a spatially uniform switch. The extraction analysis accounts for propagation through the finite pumped region and transmission into the unpumped region.

As an idealized reference, a perfectly conducting boundary is switched on instantaneously at $y=d$. Matching the interior electromagnetic fields across the switch projects the incident mode onto time-transmitted and time-reflected TE modes and a static magnetic field~\cite{Stefanini2022}. Energy-balance checks include the exterior field (Supplemental Material, SM). For incident frequency $0.35$~THz and widths $w_x=4d$, $w_z=5d$, the time-reflected TE$_1$ packet for $\ell=+1$ carries $1.41\%$ of the initial energy inside the slab and has $\Delta x_{1,+1}^{\rm R}=-42~\mu$m. The transmitted TE$_1$ packet carries more energy but shifts by only $+7.7~\mu$m because its logarithmic spectral slope is smaller (SM).
\begin{figure}[!tbp]
\includegraphics[width=\columnwidth]{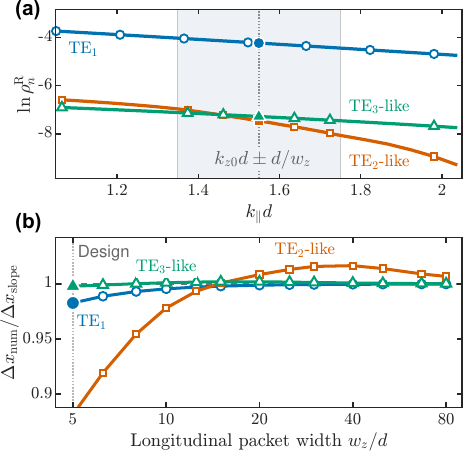}
\caption{Finite-bandwidth test of the slope law. (a)~Logarithms of the reflected spectral weights for bound TE$_1$ and the TE$_2$- and TE$_3$-like radiative bands. The vertical dotted line marks $k_{\parallel}=k_{z0}$. Shading marks $k_{z0}d\pm d/w_z$ at $w_z=5d$. (b)~Ratio $\Delta x_{\rm num}/\Delta x_{\rm slope}$, where $\Delta x_{\rm num}$ is the displacement evaluated numerically from Eq.~(\ref{eq:exact}) and $\Delta x_{\rm slope}$ is the prediction of Eq.~(\ref{eq:slope}). Filled symbols mark the design width $w_z=5d$. The horizontal dashed line indicates unity. All three ratios approach unity as $w_z$ increases.}
\label{fig:theorem}
\end{figure}

To include the finite pump duration and conductivity of the photoexcited layer, the Maxwell equations are evolved with a Drude current~\cite{Mics2013,Beard2000}. The carrier density grows during the Gaussian pump and decays exponentially with depth below the illuminated face. The calculation includes collisional loss and an explicitly meshed air exterior, with energy-balance checks given in the SM. At the reported pump fluence $1.8~\mu$J~mm$^{-2}$ and duration $100$~fs~\cite{Miyamaru2021}, the model predicts $1.4\%$ time-reflected TE$_1$ energy and $|\Delta x_{1,\ell}^{\rm R}|=41~\mu$m [Figs.~\ref{fig:concept}(c)--\ref{fig:concept}(e)]. Pump fluence and duration tune the tradeoff between conversion efficiency and displacement (Appendix~A).

Figure~\ref{fig:theorem} tests the slope law across the bound TE$_1$ channel and the TE$_2$- and TE$_3$-like frequency bands in the radiation continuum. At $w_z=5d$, the centroid displacements calculated from Eq.~(\ref{eq:exact}) are $98\%$, $88\%$, and $100\%$ of the values predicted by Eq.~(\ref{eq:slope}), respectively. For the TE$_2$-like radiative band, the deviation from the prediction of Eq.~(\ref{eq:slope}) reflects curvature of $\ln\rho$ across the packet bandwidth. All three ratios approach unity as the packet narrows spectrally. Only TE$_1$ remains guided. The subsequent extraction analysis focuses on the time-reflected TE$_1$ packet, with $\rho\equiv\rho_1^{\rm R}$ and $\Delta x_\ell\equiv\Delta x_{1,\ell}^{\rm R}$.

\textit{Extraction from the pumped region.—}The TE$_1$ extraction model combines temporal conversion, propagation loss, and transmission into the unpumped region (SM). The reflected amplitude decays as $e^{-\Gamma t}$, with $\Gamma=0.012\,c/d$ at the chosen pump parameters, where $c$ is the vacuum speed of light. The reflected energy falls to $1/e$ of the initial value after approximately $0.5$~mm of propagation in the pumped region. To limit attenuation before extraction, the pump is confined to $L_{\rm p}\simeq1$~mm, with each launched packet centered in the pumped region at the time switch. The packet center is then about $L_{\rm p}/2\simeq0.5$~mm from the exit at $z=0$. Longer propagation through the pumped region also reweights the converted spectrum because $\Gamma$ depends on $k_{\parallel}$, reducing the initial displacement.

The extracted time-reflected energy is $0.34\%$ of the incident energy (SM). The transmitted branch exits at $z=L_{\rm p}$. Both branches emerge at $0.471$~THz, spectrally separated from the unconverted $0.35$-THz component. The predicted frequency is close to the measured $0.47$~THz forward-converted frequency~\cite{Miyamaru2021}. Differences in geometry and output coupling prevent a direct comparison between the calculated and measured energies of the converted packets (SM).

\begin{figure*}[!tp]
\includegraphics[width=\textwidth]{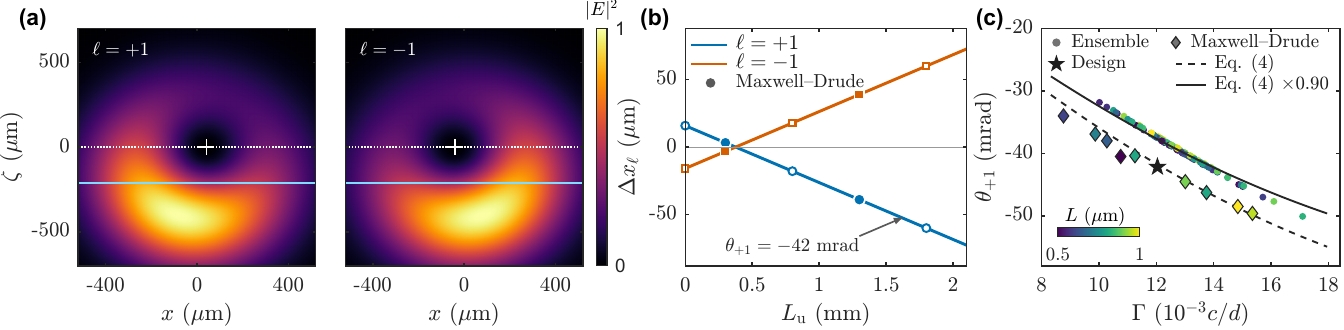}
\caption{Loss-induced deflection. (a)~Illustrative, peak-normalized TE$_1$ intensities after radial spectral filtering and exponential attenuation. Cyan lines mark longitudinal intensity centroids within the displayed window. White crosses mark phase cores at complex-field zeros, aligned with the white dotted lines. The trailing side $\zeta>0$ experiences greater attenuation. The $\ell=-1$ distribution mirrors the separately launched $\ell=+1$ distribution in $x$. (b)~Centroid half-difference $\Delta x_\ell$ versus $L_{\rm u}$ for separate launches. Filled symbols give independent Maxwell--Drude results at $L_{\rm u}=0.3$ and $1.3$~mm. Solid lines have slopes $\mp42~\mathrm{mrad}$ for $\ell=\pm1$. Open symbols mark positions on the lines. (c)~Deflection predicted by the TE$_1$ extraction model versus decay rate for $96$ material combinations at $L_{\rm p}=1$~mm, colored by absorption depth $L$. Dashed and solid lines show Eq.~(\ref{eq:kappa}) without and with the factor $0.90$ accounting for the finite pumped length. The star denotes the design-point simulation in (b), and diamonds denote nine additional Maxwell--Drude simulations at $L_{\rm p}=1.2$~mm.}
\label{fig:angle}
\end{figure*}

\textit{Angular splitting.—}An angular response requires an imbalance of transverse momentum. Energy removal can transfer transverse OAM to STOV packets~\cite{Hancock2024}. Loss-induced core drift has been studied in bulk media~\cite{Liu2026}. Portions of the time-reflected packet created deeper in the pumped region travel farther before reaching $z=0$ and experience greater loss. The extracted envelope is thus weaker on the trailing side [Fig.~\ref{fig:angle}(a)]. Unequal attenuation displaces the intensity envelope from the vortex core, preventing cancellation of the transverse momentum associated with the circulating phase. For a Gaussian vortex envelope, the derivation assumes nearly constant $\Gamma$ across the bandwidth and weak dispersive changes during extraction. Attenuation by $e^{-\Gamma t}$ then gives
\begin{equation}
\Delta\langle k_x\rangle_{\ell}=\frac{\kappa}{1+\kappa^2 w_x^2/2},
\qquad
\kappa=\frac{\ell\Gamma w_z}{v_{\rm g}w_x},
\label{eq:kappa}
\end{equation}
where $\Delta\langle k_x\rangle_\ell$ denotes the loss-induced change in the mean transverse wave number and $v_{\rm g}>0$ is the group velocity of the reflected TE$_1$ channel in the pumped region (SM). Loss generates transverse momentum inside the pumped region. With the time-reflected branch written as $e^{i(\bm k\cdot\bm r+\omega t)}$ for positive angular frequency $\omega$, the group velocity is antiparallel to \(\bm k\). After extraction, the propagation angle therefore satisfies $\theta_\ell\simeq-\Delta\langle k_x\rangle_\ell/k_z$. The longitudinal wave number $k_z>0$ is evaluated in the unpumped region at the central frequency of extracted TE$_1$ packet. For the extracted packet, the centroid half-difference varies with propagation distance as $\Delta x_\ell(L_{\rm u})=\Delta x_\ell(0)+\theta_\ell L_{\rm u}$. A linear fit yields $\theta_\ell$ from the slope and the exit-plane offset $\Delta x_\ell(0)$ from the intercept [Fig.~\ref{fig:angle}(b)]. The fitted offset includes centroid shifts introduced during extraction. The relation between offset and displacement at the time boundary also depends on how the field is recorded.

Electro-optic sampling reconstructs $E(x,t)$ at a fixed detection plane. The transverse centroid follows from integrating $|E(x,t)|^2$ over time, whereas Eq.~(\ref{eq:exact}) gives the centroid at a fixed time. The guided dispersion maps the longitudinal wave number $k_z$ to the angular frequency $\omega$ at fixed $k_x$, shearing the recorded vortex (SM). For $\ell=+1$, the spatial-to-temporal mapping contributes approximately $\ell w_z/(2w_xk_{z0})=+40~\mu$m, nearly canceling the $-41~\mu$m conversion displacement. The centroid measured at a single plane therefore does not directly give the displacement produced at the time boundary.

A two-dimensional Maxwell--Drude solver reconstructs separately launched $\ell=\pm1$ packets from $|k_x|\le1/d$ samples for $L_{\rm p}=1.2$~mm. The coupled Maxwell and Drude equations directly determine temporal conversion and attenuation. Before accounting for the detector aperture and response, the centroid displacement is $\Delta x_{+1}=+3~\mu$m at $L_{\rm u}=0.3$~mm and $-39~\mu$m at $L_{\rm u}=1.3$~mm. The two-plane drift and the mean transverse wave number are two estimators applied to this same Maxwell--Drude field. Both give $\theta_{+1}\simeq-42~\mathrm{mrad}$ at the quoted precision. The separately constructed TE$_1$ extraction model reproduces the angular response for the matched input preparation (SM).

A phenomenological correction near the ends of pumped region reduces the centroid difference between the extraction model and Maxwell--Drude calculation (Appendix~B). The spatial profile and characteristic length of this correction affect the absolute centroid position. The transverse momentum is less sensitive to correction in the tested geometry, so agreement of the angular response does not fix the absolute position offset.

\textit{Detection and parameter dependence.—}Time-delayed electro-optic sampling can reconstruct terahertz STOVs~\cite{Wang2023}. In the proposed measurement, the extracted centroid is recorded at several propagation distances $L_{\rm u}$ for separately launched $\ell=\pm1$ packets. Each charge has its own centroid trace, which is fitted independently. The model predicts equal and opposite slopes. The fits do not impose this relation. The sum of the fitted slopes provides a null test for charge-independent drift. For $L_{\rm p}=1.2$~mm and $\ell=+1$, detector filtering changes the calculated centroid slope from $-42$ to $-23.9~\mu$m/mm (Appendix~D). Aperture weighting reduces the magnitude but preserves charge reversal and approximate linearity over the modeled range.

Under the noise assumptions in Appendix~D, Monte Carlo simulations give a total acquisition time of $6.7$~min for both charges at six planes. For each charge, the estimated root-mean-square error of the slope is below $20\%$ of the predicted magnitude. This time excludes movement and calibration. For $\ell=+1$, the conversion-induced displacement and the deflection angle remain negative across the $96$ tested combinations of pump and carrier parameters. The angle $\theta_{+1}$ ranges from $-50$ to $-32~\mathrm{mrad}$ and varies with the decay rate $\Gamma$ [Fig.~\ref{fig:angle}(c), Appendix~C]. Nine additional Maxwell--Drude calculations test the loss dependence beyond the design point.

\textit{Conclusion.—}Radial spectral filtering during temporal conversion couples STOV phase circulation to transverse position. For the specified scalar spectrum, the narrowband displacement depends on the logarithmic slope of $\rho_n^p$ and the packet aspect ratio. Radial spectral filtering alone leaves the mean transverse momentum zero. For the specified scalar Gaussian input, the displacement law provides a criterion for platforms with a radial conversion coefficient and a smooth, nonzero spectral weight over the narrow packet bandwidth. In the GaAs realization, unequal attenuation during extraction supplies transverse momentum and produces opposite propagation angles, as tested by independent Maxwell--Drude simulations. The angle after extraction depends on the loss, dispersion, and geometry of each platform. Charge reversal and measurements at multiple planes distinguish displacement from deflection.

\paragraph{Acknowledgments.} The research was supported by the National Natural Science Foundation of China (62275015 and 62205015).

\paragraph{Data and code availability.} The data that support the findings of this article,  simulation and plotting code are available from the authors upon reasonable request.

\bibliography{references}

\onecolumngrid
\vspace{4pt}
\begin{center}
\rule{0.45\textwidth}{0.4pt}\\[3pt]
\textbf{End Matter}
\end{center}
\vspace{1pt}
\twocolumngrid
\makeatletter
\@booleantrue\raggedcolumn@sw
\makeatother
\raggedbottom

\section{Appendix A. Pump control of the conversion spectrum}

\begin{figure}[!hb]
\includegraphics[width=\columnwidth]{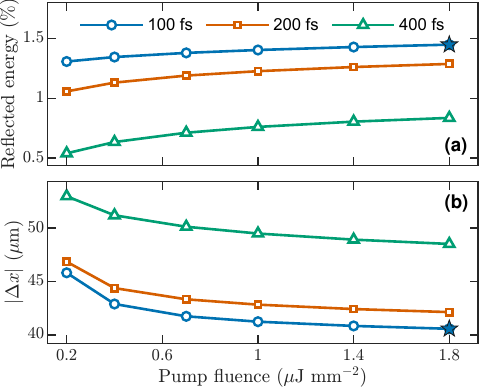}
\caption{Pump control at absorption depth $L=1~\mu$m. (a)~Reflected TE$_1$ energy as a percentage of the initial energy inside the slab and (b)~displacement magnitude at the time boundary versus pump fluence. Blue circles, orange squares, and green triangles denote pump durations of $100$, $200$, and $400$~fs. The stars mark calculated values at the reported pump fluence of $1.8~\mu$J~mm$^{-2}$ and duration of $100$~fs~\cite{Miyamaru2021}.}
\label{fig:pump}
\end{figure}

The instantaneous boundary provides a nonadiabatic reference~\cite{Galiffi2022}. Figure~\ref{fig:pump} shows how pump fluence and duration control reflected TE$_1$ energy and displacement for absorption depth $L=1~\mu$m. At fixed duration, increasing fluence raises the converted energy but flattens $\ln\rho$, reducing the displacement. At fixed fluence, increasing duration lowers the energy and steepens $\ln\rho$. The reported fluence $1.8~\mu$J~mm$^{-2}$ and duration $100$~fs~\cite{Miyamaru2021} give a modeled reflected energy fraction of $1.4\%$ and displacement magnitude of $41~\mu$m.

\section{Appendix B. Correction near the ends of the pumped region}

In the independent time-domain benchmark, the incident spectrum is specified through TE-mode amplitudes of the vector potential before energy normalization. For a consistent comparison, the TE$_1$ extraction model retains the modal energy weight associated with this input preparation (SM).

The pumped and unpumped regions meet at $z=0$ and $z=L_{\rm p}$. The TE$_1$ extraction model applies the conversion weight of a uniformly pumped waveguide to the finite region. For $L_{\rm p}=1.2$~mm, $|k_x|\le0.5/d$, and $L_{\rm u}=0.3$~mm, the TE$_1$ extraction model gives $\Delta x_{+1}=-12.5~\mu$m, compared with $+5.1~\mu$m from the Maxwell--Drude simulation. At the design carrier, $\tilde{k}_zd=1.555+0.101i$ gives a guided wavelength $4.04d$ and an intensity attenuation length $4.93d$. A phenomenological interpolation reduces the logarithmic spectral slope near the two ends. When the interpolation length is set to the guided wavelength, the centroid difference decreases from $17.6~\mu$m to $1.7~\mu$m. The predicted mean transverse wave number changes from $+0.099/d$ to $+0.095/d$. The Maxwell--Drude value is $+0.102/d$. Setting the interpolation length to the attenuation length leaves a $4.7~\mu$m centroid difference. The positional offset depends on the interpolation profile and length, which remain phenomenological model choices.

\begin{figure}[!tbp]
\includegraphics[width=\columnwidth]{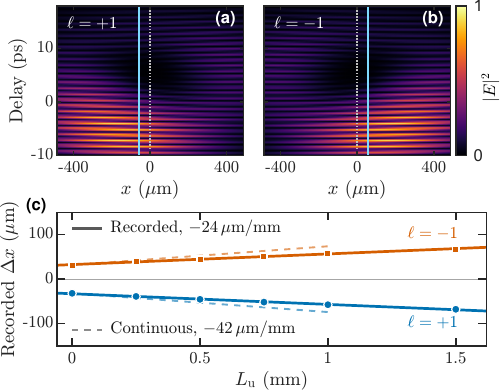}
\caption{Electro-optic record predicted by the TE$_1$ extraction model for $L_{\rm p}=1.2$~mm. (a),(b)~Peak-normalized $E_{\rm EO}(x,t)^2$ of the real electro-optic field for separately launched $\ell=\pm1$ packets at $L_{\rm u}=1$~mm, through a $1$-mm aperture and a $3$-mm-thick ZnTe detector. Delay is measured from the temporal centroid of the real-field intensity. Dotted and cyan solid lines mark $x=0$ and the transverse centroids $\mp57~\mu$m, respectively. Centroids use trapezoidal integration over all $321$ delays. (c)~Symbols show calculated centroids versus $L_{\rm u}$. Solid lines are linear fits with slopes of $\mp24~\mu$m/mm for $\ell=\pm1$. Dashed lines have slopes of $\mp42~\mu$m/mm before detector filtering and pass through the recorded centroids at $L_{\rm u}=0$.}
\label{fig:detector}
\end{figure}

\section{Appendix C. Robustness over the material ensemble}

The ensemble samples $96$ deterministic combinations of pump fluence, duration, reflectivity, carrier depth, yield, and collision time with a fixed Latin-hypercube design over the bounds listed in the SM. Within these sampled bounds, all combinations retain negative displacement at the switch and negative drift for $\ell=+1$. The conservative reflected energy at the switch spans $1.32$--$1.42\%$. This estimate excludes the conversion contribution from $k_{\parallel}d<1$ while retaining the full incident-energy normalization. Extracted energy at $L_{\rm p}=1$~mm spans $0.24$--$0.38\%$. The angle ranges from $-50$ to $-32~\mathrm{mrad}$, with Spearman rank correlation $0.999$ between drift magnitude and $\Gamma$ [Fig.~\ref{fig:angle}(c)]. Nine further Maxwell--Drude calculations at $L_{\rm p}=1.2$~mm follow the same loss dependence. Spatial truncation reduces the predicted angle less at $L_{\rm p}=1.2$~mm than at $L_{\rm p}=1$~mm. The reported intervals are extrema of the sampled parameter set, not confidence intervals.

\section{Appendix D. What the sampler records}

The centroid is evaluated after applying the detector aperture and transfer function to the field. For Fig.~\ref{fig:detector}, the exit field of the $L_{\rm p}=1.2$~mm region is first propagated through the unpumped region. The field is then filtered by a $1$-mm aperture, the $3$~mm ZnTe response, and the spatial point-spread function~\cite{Gallot1999}. The real electro-optic field is sampled on the $40$-pixel, $321$-delay grid specified in the SM. Its squared field is integrated over delay to obtain the centroid. Doubling the delay resolution changes any centroid by less than $0.006~\mu$m. The recorded centroid reverses with charge. Aperture weighting emphasizes the core and suppresses contributions from the wings, reducing the drift from approximately $-42$ to $-24~\mu$m/mm for $\ell=+1$. The calculated centroids deviate from a linear fit in $L_{\rm u}$ by at most $1.02~\mu$m over the sampled range. Thus the detector changes the measured magnitude while preserving the reversal of the propagation angle with topological charge.

The centroid half-difference at each plane is formed from the two separately recorded extracted packets, without subtracting the centroid of an incident reference. For an incident peak field of $100$~V~cm$^{-1}$, the detector model gives a peak field near $5.9$~V~cm$^{-1}$ (SM).

Monte Carlo simulations of real electro-optic field records estimate the averaging requirements under the noise model specified in the SM~\cite{Ibrahim2016,Shields2022}. For the reference-camera configuration, $182$ averages at each of $321$ delays require $6.7$~min for both charges at six planes. This time excludes movement between planes and calibration. The simulations include residual errors in the transverse coordinate and residual background contributions. These contributions remain fixed during an acquisition, so averaging does not reduce them.

The same number of averages is tested with independent noise samples. For each charge, the estimated root-mean-square error of the slope remains below $20\%$ of the predicted slope magnitude (SM).

A systematic error in the transverse coordinate can also produce an apparent centroid drift. If the error varies linearly with propagation distance, it changes the fitted slope. Equal and opposite slope errors for the two charges cancel in the sum of the fitted slopes. Independent coordinate calibration is therefore needed to distinguish such measurement errors from the predicted deflection (SM).

\end{document}